\def\papertitle{A Generative Model For Raw Audio Using Transformer Architectures}
\def\paperauthorA{Prateek Verma}
\def\paperauthorB{Chris Chafe}

\documentclass[twoside,a4paper]{article}
\usepackage{etoolbox}
\usepackage{dafx_20}
\usepackage{amsmath,amssymb,amsfonts,amsthm}
\usepackage{euscript}
\usepackage[T1]{fontenc}
\usepackage[utf8]{inputenc}
\usepackage{nimbusserif}
\usepackage{ifpdf}
\usepackage[english]{babel}
\usepackage{caption}
\usepackage{subfig} 
\usepackage{color}

\input glyphtounicode
\ninept

\newcounter{numauth}
\newcounter{listcnt}
\newcommand\authcnt[1]{\ifdefined#1 \stepcounter{numauth} \fi}

\newcommand\addauth[1]{
\ifdefined#1 
\stepcounter{listcnt}
\ifnum \value{listcnt}<\value{numauth}
\appto\authorslist{, #1}
\else
\appto\authorslist{~and~#1}
\fi
\fi}
\authcnt{\paperauthorB}
\def\authorslist{\paperauthorA}
\addauth{\paperauthorB}
\addauth{\paperauthorC}
\addauth{\paperauthorD}
\addauth{\paperauthorE}
\addauth{\paperauthorF}
\addauth{\paperauthorG}
\addauth{\paperauthorH}
\addauth{\paperauthorI}
\addauth{\paperauthorJ}

\usepackage{times}

\newif\ifpdf
\ifx\pdfoutput\relax
\else
   \ifcase\pdfoutput
      \pdffalse
   \else
      \pdftrue
\fi

\ifpdf 
  \usepackage[pdftex,
    pdftitle={\papertitle},
    pdfauthor={\authorslist},
    pdfsubject={Proceedings of the 23rd International Conference on Digital Audio Effects (DAFx-20)},
    colorlinks=false, 
    bookmarksnumbered, 
    pdfstartview=XYZ 
  ]{hyperref}
  \usepackage[pdftex]{graphicx}
\else 
  \usepackage[dvips]{epsfig,graphicx}
  \usepackage[dvips,
    pdftitle={\papertitle},
    pdfauthor={\authorslist},
    pdfsubject={Proceedings of the 23rd International Conference on Digital Audio Effects (DAFx-20)},
    colorlinks=false, 
    bookmarksnumbered, 
    pdfstartview=XYZ 
  ]{hyperref}
\fi
\usepackage[hypcap=true]{caption}
\title{\papertitle}

\affiliation{
\paperauthorA\,\sthanks{Thanks to the predecessors for the templates}and \paperauthorB \,\sthanks{This work was supported by the XYZ Foundation}}
{\href{https://www.mdw.ac.at/ike/}{Institute 1} \\ University of Music and Performing Arts\\ Vienna, Austria\\
{\tt \href{mailto:dafx2020@gmail.com}{dafx2020@gmail.com}}
}

\affiliation{
\paperauthorA\ and \paperauthorB\,\thanks{\vspace{-3mm}}}
{\href{https://www.mdw.ac.at/ike/}{Center for Computer Research in Music and Acoustics} \\ Stanford University, Stanford, CA, USA\\
{\tt \href{mailto:prateekv@stanford.edu}{prateekv@stanford.edu} | \href{mailto:cc@ccrma.stanford.edu}{cc@ccrma.stanford.edu} }
}
\usepackage{cuted}
\usepackage{capt-of}
\begin{document}
\ifpdf 
  \DeclareGraphicsExtensions{.png,.jpg,.pdf}
\else  
  \DeclareGraphicsExtensions{.eps}
\fi


\maketitle

\begin{abstract}
This paper proposes a novel way of doing audio synthesis at the waveform level using Transformer architectures. We propose a deep neural network for generating waveforms, similar to wavenet \cite{oord2016wavenet}. This is fully probabilistic, auto-regressive, and causal, i.e. each sample generated depends on only the previously observed samples. Our approach outperforms a widely used wavenet architecture by up to 9\% on a similar dataset for predicting the next step. Using the attention mechanism, we enable the architecture to learn which audio samples are important for the prediction of the future sample. We show how causal transformer generative models can be used for raw waveform synthesis. We also show that this performance can be improved by another 2\% by conditioning samples over a wider context. The flexibility of the current model to synthesize audio from latent representations suggests a large number of potential applications. The novel approach of using generative transformer architectures for raw audio synthesis is, however, still far away from generating any meaningful music similar to wavenet, without using latent codes/meta-data to aid the generation process. 
\end{abstract}

\section{Introduction and Related Work}
\begin{sloppypar}
Audio synthesis has long fascinated musicians, computer scientists, and researchers and has been one of the core building blocks of computer music. FM synthesis by John Chowning \cite{chowning1973synthesis} is an example of distortion-based frequency domain synthesis with which complex, time-varying spectra can be modeled. An early time-domain technique that generated a particular class of realistic audio was the Karplus Strong algorithm \cite{karplus1983digital}. An early example presaging physical modeling, it could generate transient waveforms using a filtered delay line loop for the synthesis of plucked string and percussion-like sounds. These are examples of techniques that generate waveforms given a set of parameters. The difficulty of modeling musical structure at longer time scales has been well summarized in \cite{dieleman2018challenge}. In this paper, we propose a generative framework using transformers to synthesize audio waveforms. We show how the approach can outperform the classic wavenet model \cite{oord2016wavenet} that has become a fundamental building block for a variety of audio synthesis frameworks. Wavenet, based upon causal dilated convolutional filters, is an auto-regressive architecture. By conditioning on the desired metadata, it surpassed state-of-the-art Text-to-Speech synthesis in 2016.  Wavenet-based sub-block architectures have been used subsequently in problems such as speech denoising \cite{rethage2018wavenet}, instrument conversion \cite{mor2018universal}, and as a vocoder \cite{shen2018natural} to go from spectral to time-domain signals. For problems such as source separation and denoising the generative process is conditioned implicitly on a learned latent representation, where for vocoder-based application, it is conditioned explicitly on a spectral representation (typically mel-spectrum) which guides the generation process. An end-to-end architecture, proposed in \cite{stoller2018wave,tamamori2017speaker} used a dilated convolution-based generator for end-to-end source separation. 
High-fidelity synthesis was proposed in \cite{oord2018parallel} using a teacher-student framework, by utilization of the idea of distillation to guide the training of neural architecture. Using self-supervised learning, where the goal is to predict the next sample given the previous context, wavenet can also learn latent representations in problems such as speech recognition. Representation learning has been an active area of research in supervised and self-supervised setups \cite{verma2019neuralogram,verma2020framework,devlin2018bert}. As we can see, there are a variety of applications encompassing a broad spectrum that employ wavenet-based generators for waveform synthesis. The main advantage of end-to-end learning is to mitigate the transformation from spectrogram to waveform for audio signals. Spectral inversion has been explored in a variety of ways, with the classic work \cite{griffin1984signal} which converts magnitude STFT spectra to time-domain signals. The idea of synthesis/generation conditioned on the external application has been explored for a variety of domains and applications. It enables the fixed-parameter neural architectures to be more expressive in terms of controlling/guiding the output. It was shown in \cite{haque2018conditional,verma2018neural, guo2019end} that by conditioning wavenet-inspired architectures, audio can be converted from one domain to another. Wavenet-inspired models have also shown success in areas such as speech recognition and latent representation learning from spectral inputs \cite{haque2019audio}. Longer-term contexts, captured by latent representation, can then guide a synthesis network for packet loss concealment approach as described in \cite{verma2020deep}. The approach has been explored in natural language processing too, to guide generated language according to metadata such as sentiment and ratings \cite{keskar2019ctrl}. Problems such as text-to-speech synthesis, explored conditioning according to desired characteristics such as speaker ID, prosody, sentiment to guide the generated spectral representation. \cite{wang2017tacotron}. 
\begin{figure*}\centering
\includegraphics[width=\textwidth,height=5.6cm]{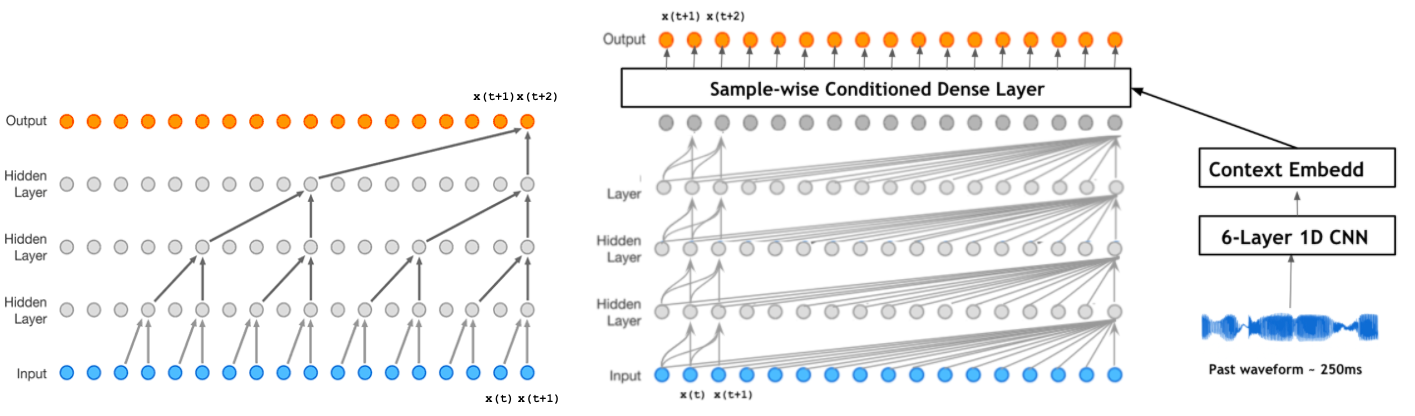}
\captionof{figure}{An overview of the proposed method in the paper. Classical wavenet architecture(left) and our proposed conditioned generative transformer architecture(right) for raw audio synthesis. 
\label{fig:feature-graphic}}
\end{figure*}
For audio analysis, early studies have shown how initial layers of neural architecture can learn spectral representations in speech recognition \cite{sainath2015learning} and frequency estimation \cite{verma2016frequency}. The front end adapts according to the problem of interest. In \cite{verma2016frequency} the front end adapts to learn a non-linear non-constant bandwidth filter-bank that outperforms traditional Fourier representations (which are fixed for all applications). This gives the network an ability to first implicitly learn an analysis-synthesis pipeline based upon internal latent representations. There have been several variants and advancements improving on shortcomings of the wavenet architecture. Researchers in \cite{kalchbrenner2018efficient} proposed an architecture that could do raw waveform auto-regressive synthesis on raw waveforms using CPU only, almost in real-time. 
\end{sloppypar}
The core idea in this paper is to apply Transformer architectures. First proposed in \cite{vaswani2017attention}, these have revolutionized modern deep learning by providing an ability for modeling long-term sequences.  The basic premise is first to learn important parts of the input via the technique's attention mechanism, followed by feed-forward architectures that can map them to a separable appropriate latent space. This can be repeated at multiple levels to learn a hierarchy of features. The idea was proposed in 1991 \cite{WinNT}. With this simplicity, it turns out these architectures can have enormous power to model a variety of modalities, such as music \cite{huang2018music}, audio \cite{verma2021audio}, text \cite{brown2020language,devlin2018bert}, protein sequences \cite{madani2020progen}, videos\cite{sun2019videobert,girdhar2019video} and vision \cite{dosovitskiy2020image,parmar2018image} to name a few. The main drawback comes with the attention block which scales quadratically both in terms of computational and memory constraints. This reduces their effectiveness at scale, such as when attempting much longer sequences, for example, 5 minutes of music at the waveform level \cite{dieleman2018challenge}. It is however an active area of research with possibilities that remain to be explored.  Imposing sparsity \cite{child2019generating} has been shown to improve the capability of these architectures over longer-term sequences. Other techniques have tried learning latent representation using VQ-VAEs \cite{oord2017neural}. The idea of learning on latent representations of the input can be used to encapsulate various attributes of the input signal in a compact code, like pitch, timbre, and rhythm for music, as shown in \cite{verma2019neuralogram}. It enables to model audio signal at a much smaller scale which the techniques used in \cite{vaswani2017attention} can easily capture. By modeling the latent representations of the future with a transformer, one can guide a conditioned waveform generator on these future latent predictions. This was done by \cite{dhariwal2020jukebox} to produce compelling results. It was, however, based on heavy conditioning, and was not an end-to-end approach. Similar approaches were used for problems such as speech recognition \cite{baevski2020wav2vec,baevski2019vq} to mimic BERT \cite{devlin2018bert} pre-training for speech signals. The main reason for the good performance of these architectures are i) The ability of these architectures to model long sequences ii) self-supervised training enables the performance to scale well with data \cite{ellis1999size}. Recent approaches of improving Transformers by combining ideas from Mixture of Expert models \cite{fedus_zoph_shazeer_2021} using up to a trillion parameters makes the current work and the road ahead even more exciting.

The contributions of the current work are as follows:\newline

i) We propose a state-of-the-art neural network architecture that outperforms the traditional wavenet architecture by 6-11\% on a similar dataset with the task of prediction of the next sample.

ii) We propose conditioned Transformer-based generative models that show that they can give a further performance boost over the unconditioned Transformer. This has the potential for a variety of setups and conditions in the future.

iii) As with the classic wavenet architecture, this provides a new method for many applications that rely on neural architectures for a conditioned audio generation like TTS, instrument/voice transformation, speech denoising, source separation, music generation, spectral inversion to name a few. 

Section I gives a brief introduction and outlines related work in the field,  Section II describes the dataset used for the current problem, followed by the methodology used in our paper. Section 4 compares the results with baseline wavenet architectures followed by the conclusion and future work in Section 6.
\section{Dataset}
In this work, we use real-world piano recordings from Youtube. The choice of the piano was made as it contains a mixture of both monophonic and polyphonic (playing more than one note at a time) thus yielding complex audio waveforms. The piano also provides a wide range of pitches. However, it does have a relatively constrained timbre space as opposed to more choices we could have made, such as symphonic performances. The flexibility of conditioning the generation on a latent code makes it adaptable to further datasets. We collected about 840 piano tracks of covers of popular songs amounting to about 56 hours of data. The reason for selecting real-world recordings over a standard dataset \cite{bittner2014medleydb,hawthorne2018enabling} was to account for differences in room acoustics, playing styles, timbral differences, loudness, and microphones to name a few. The tracks included artists ranging from Pink Floyd and Billie-Eilish to some works from Rahman and Coke-Studio among many others. Forty randomly chosen tracks were set aside as a test set for which all the results in the paper are reported. The URLs are shared here at https://tinyurl.com/6cjxffa8

For training, we worked with a 16KHz sampling rate and 8-bit resolution. All the tracks were read and converted to the chosen resolution using \cite{virtanen2020scipy,mcfee2015librosa}. Any tracks having lower original sampling rates were discarded.  The context duration was 100ms, a choice required by computation cost and memory constraints for the resources available, and because of the quadratic scaling of transformers. The same setup and context were used for the wavenet comparisons, too. Since this is a form of self-supervised learning where the data itself is used, the input vocabulary consisted of the raw waveform representation, values ranging from 0-255. The output target was the same waveform but shifted by one so that at each time step we predict the next sample. This is similar to a recent language modeling project \cite{brown2020language}. All of the models discussed in the current paper are causal attention dilation-based. Given that we used such a large dataset, we expect the results and improvement of Transformer over wavenet to hold on to a standard dataset. \footnote{We have shared our code for models, the youtube-URLs, and the training and testing tracks here https://tinyurl.com/6cjxffa8
}
\section{Methodology}

This work proposes a generative model for audio synthesis. This is done by directly modeling the probability distribution of the samples of the waveform of an audio signal. The audio signal represents changes in the amplitude of a signal over a certain time as sound waves traverse through space. There are two main characteristics of the waveform beyond the amplitude, when we choose to digitize the signal, i.e the sampling rate which measures how often we sample the continuous signal, and the bit-resolution, i.e. how many bits we allocate to represent the signal in the range of -1 to 1. For our current work, we choose to have an 8-bit resolution (similar to classic wavenet) and 16kHz. Raw audio synthesis at higher bit rates becomes a difficult problem due to the sheer amount of states involved (65536 and ~ 16M states respectively for 16-bit and 24-bit audio signals) and is the subject of an ongoing current work.

We model the probability of a waveform as follows. Let $\textbf{x}$  denote the joint probability of any observed waveform of length $T$, as described in \cite{oord2016wavenet}. The joint probability of a waveform, $\textbf{x} = {x_1, x_2, .... x_{T}}$ is modelled as,

\begin{equation}
    p(\textbf{x}) = \prod_{t=1}^{T} p(x_{t}|x_1, x_2, .... x_{t-1})
\end{equation}


\subsection{Baseline: Wavenet architecture} To compare our method with classic neural network-based autoregressive wavenet architecture, we describe here the two variants that were implemented and compared against. This is done mainly due to the lack of significant implementation details in the original paper \cite{oord2016wavenet}. The main ingredients of the paper are causal dilated convolutions, which ensures that at each time step, the model only looks into the past samples to generate the future sample. By using dilated convolutions, we enable the model to learn longer contexts through large receptive fields. The figure below shows, the architecture of dilated convolutional model for waveform generation taken from \cite{oord2016wavenet}.

\begin{figure}[ht]
\centerline{\includegraphics[scale=0.31]{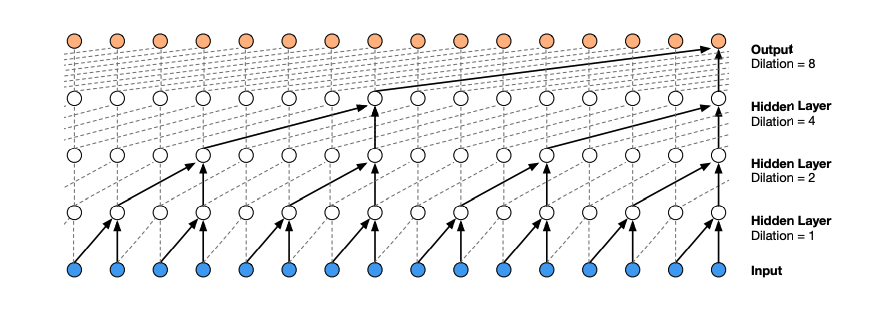}}
\caption{\label{wavenet}{\it Wavenet architecture showing how dilated convolutions can model long term dependencies. The figure is from \cite{oord2016wavenet}}}
\end{figure}
As seen from Figure 2, the architecture consists of increasing the receptive field size at higher layers by dilating (skipping) the convolution at each layer by a fixed factor. At the very outset, we can see the advantage attention-based architectures might have over the classic wavenet architecture as described in Figure 1. A Transformer model iterates over all the possible connections that can happen with the context and chooses which ones are important over others as opposed to fixed connections that look over a pre-defined topology in wavenet. By using multiple layers of Transformer modules the networks learn to also learn higher-order dependencies across the learned representations in the previous layer. This is explained in more detail in the next section.

\subsection{Wavenet Implementation details}
In our work, we increase the dilation rate by a factor of 2 in each layer. In our baseline model, we had a total of 10 layers, with 128 convolutional filters in each layer thus yielding an effective receptive field size of 1024 in the final layer.  Similar to the wavenet paper,  our second model consists of stacking three such layers of causal dilated convolutions, with the filter size of 2, dilated by a factor of 2 in each of the subsequent layers. This in total resulted in about 30 layers of causal convolutions with dilation rate repeated as 1,2,4, .... 512, 1,2,4....512, 1,2,4,...512. This is a very strong baseline system to compare. This results in more parameters and an increase in the receptive field size and model capacity. Skip connections were used for better convergence, speedup, and the ability of the network to learn residual inputs as it has shown to improve across images, text, and audio. A dense layer is used to convert the output of the final layer to that of a fixed 256-dimensional encoding. A soft-max layer converted the logits to a probability distribution, and cross-entropy loss was minimized between the ground-truth bit level and the predicted one. This setup is maintained for all of the models discussed in this work. A reason for selecting softmax over traditional other error criteria such as Euclidean Loss, is due to i) as described in \cite{oord2016wavenet}  "categorical distribution" is more flexible can more easily model arbitrary distributions because it makes no assumptions about their shape." and ii) categorical distributions impose much stricter penalty than Euclidean distance as the error penalty is the same if the predictions are even off by a single bit level than say by larger numbers. 
This is also a type of self-supervised architecture in the sense that data itself is used for training. For all of the current models, the context chosen was 100ms, i.e. 1600 samples. This was primarily because of the huge memory costs that Transformer Architectures fail to mitigate for longer context, and the constraints of the GPUs. Additionally, the original wavenet architecture had a context of 240ms, which is twice as large, but all of the current results are reported on the same dataset, with the same context of 100ms for all the three architectures in a similar experimental setup. 

\subsection{Generative Transformer Architecture over Waveforms} 

\begin{figure}[ht]
\centerline{\includegraphics[scale=0.38]{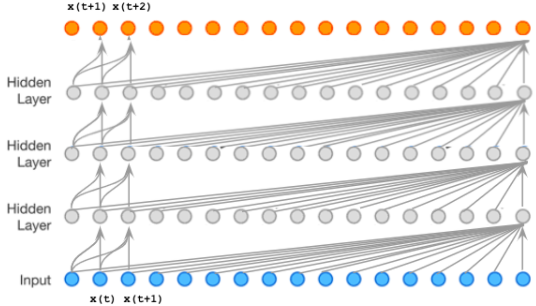}}
\caption{\label{generative}{\it Generative Transformer over Raw waveforms. Notice how the attention mechanism can learn which parts of the inputs are important at every layer, as opposed to fixed topology of the dilated convolutional based approaches.}}
\end{figure}

This section describes the Generative Transformer Architecture as described in \cite{vaswani2017attention} that we used to train the system as shown in Figure 3. A detailed explanation is given in \cite{opennmt}, but for the sake of clarity and completeness, we describe it here. As a black-box, which we would describe in more detail in this section, it takes as an input vector sequence of a fixed length $T$ during training, and produces the same length but with a chosen dimension, which we call $E$, which denotes the size of the latent space. More specifically,  it maps a sequence of vectors $\textbf{x} = {(x_1, x_2, .... x_{T})}$ to a sequence of vectors of same length $T$,  namely $\textbf{z}: ({z_1, z_2, .... z_{T}})$, where each of the dimensions of  $({z_1, z_2, .... z_{T}})$ is the chosen hyper-parameter $E$, which in our case is 128, the size of the embedding. For the sake of brevity, we would explain only one Transformer Encoder, and for a model with layers, $L$, each of the stacks is super-imposed on the other one.

Each Transformer module consists of an attention block and a feed-forward block. The output of each of them is passed through a layer norm and a residual layer. So after both the attention block and the feed-forward block, if the input to a sub-block (attention $\mathrm{F}_a$ or feed-forward $\mathrm{F}_{ff}$ block) is a sequence $x_b$, instead of passing the output directly  to the next module/sub-block, we pass along the block layer norm and the residual output $x_{bo}$
$$ x_{bo}=\mathrm{LayerNorm} (x_b + \mathrm{F}_{a/ff}(x_b))$$
This follows the notion that layer-norm/skip connections help in better convergence/improved performance. 
We now describe each of the two sub-blocks that are part of the transformer block namely, i) multi-headed causal attention ii) feed-forward architecture

\subsubsection{Multi-Headed Causal Attention} In layman's terms a multi-headed causal attention function can be described as a weighting function that decides how to get the output of each step. We weigh the rest of the inputs that are fed onto it. In other words, it assigns a probabilistic score of how important each of the embeddings is while predicting the output.  Multi-headed attention consists of first learning a probabilistic score. It is then multiplied with each of the inputs to determine how important each of the inputs is for the prediction of the embedding for a position $pos$ belonging to  $1,2,3....T$. The attention mechanism used in our case is scaled-dot product attention. With each of the inputs at every position, we learn a query, key, and a value vector. This is done by implicitly learning matrices, $W_Q$, $W_K$, and $W_V$ to produce a query vector $q$, key vector $k$, and value vector $v$ for each of the inputs for a single attention head.  We take the dot product of query and key vectors, the result is multiplied by a normalization factor (the inverse of the square root of the size of the vector as done in \cite{vaswani2017attention}), before taking a soft-max across all the inputs. Each of the value vectors is multiplied by this score to get the output of the attention module. Mathematically, for a query matrix $Q$, key matrix $K$, and a value matrix $V$, it is defined as,

$$ \mathrm{Attention}(Q,K,V) = \mathrm{softmax}(\frac{QK^T}{\sqrt{d_k}})$$

We can also learn multiple such attention maps for $h$ attention heads, defined as,

$$\mathrm{MultiHeadAttention}(Q,K,V) = \mathrm{Concat}(h_1,h_2,...h_h)W_o $$, where each of the attention heads $h_i$ is defined as 
$$\mathrm{Attention}(Q_i,K_i,V_i) = \mathrm{softmax}(\frac{Q_iK_i^T}{\sqrt{d_k}})$$
and $W_o$ is a matrix learned during training. In this work, we focus on a causal attention map which is made possible by multiplying with a mask of a triangular matrix so that each of the attention head only gives weightage to the previous sample at position $pos$ and all the future entries are set to zero. This is critical as our output is the input signal shifted by one as the norm is for training language models. \cite{brown2020language}

\subsubsection {Additional Architecture Details} Once we have weighted the importance of each of the input via multi-headed attention for passing at a position $pos$, we pass the input embedding at each position $pos$ through a feed-forward architecture. For the dimension of feed-forward layers $d_{ff}$, we have the output of the feed-forward layers $x_{bo}$ for an input $x_b$ for a 2-layer network as,

$$ \mathrm{FF}(x_b) = \max(0,x_bW_{1}+ b_{1})W_{2} + b_2. $$

This function is identically applied to each position of the input. As described in \cite{vaswani2017attention}, we add positional encodings to the input.  The model does not take into account the relative positions of the waveform and the input is passed on as a list. For any position $pos$ for the dimension $i$ of the latent space, we use a sinusoidal function, i.e. to each position $pos$ and embedding dimension $i$ in $E$, we add, $$ \mathrm{PE}_{pos,2i} = \sin(pos/10000^{(2i/E)})$$
$$\mathrm{PE}_{pos,2i+1} = \cos(pos/10000^{(2i/E)})$$

This adds positional information for each point in time and embedding dimension E as input before passing thorough self-attention layers. The output of the last Transformer block is then passed through a dense layer, typically representative of the size of the output space in our case 8-bit waveform levels.  Finally, the logits are converted to probability distribution using softmax activation. We also minimize cross-entropy loss as mentioned in the previous section, and also to be consistent with wavenet training.

\subsection{Conditional Generative Transformer}

\begin{figure}[ht]
\centerline{\includegraphics[scale=0.2]{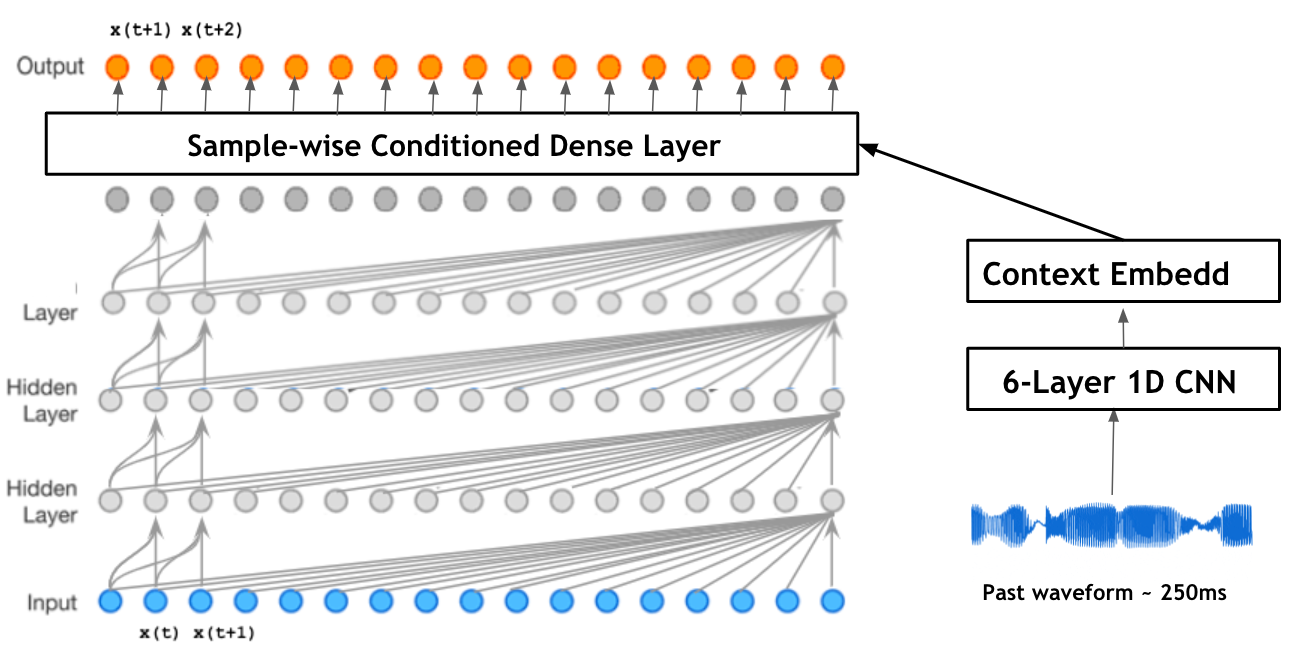}}
\caption{\label{transformer}{\it Our proposed conditioned transformer architecture. This improves performance, and helps mitigate the quadratic memory constraint of the classical transformers marginally }}
\end{figure}

Conditional generative transformers have been shown to guide the generative output according to the desired attributes. They have been deployed in areas such as NLP \cite{keskar2019ctrl}, to condition the generation of text according to sentiment, bias to reviews, or any other meta-data we choose to have. In our current work, to give more context to the generation process, we condition it with the previous context as shown in Figure 4. Due to the quadratic scaling of the attention mechanism, it is difficult to attend to long-term sequences, and thus scale this to higher context sizes say 4000 samples or beyond. To circumvent this, we use a convolution architecture to learn a latent space, which might be suitable. We achieve this by using a 6-layer convolutional architecture on the past 250ms, i.e. 4000 samples, each having 128 filters and down-sampling by 2, followed by a dense layer to learn a latent space of size 128. There are a variety of ways we can condition an implicitly or explicitly learned latent space. We follow the late fusion approach. Our model combines the output logits and the latent space learned through a convolutional net through a dense layer, and every output prediction is conditioned on the latent space.  

\subsection{Unconditioned Raw Waveform Generation}

We explored, similar to char-rnn models if we can generate new audio purely by sampling from the trained models iteratively. To synthesize new audio, since the model is probabilistic, we feed just the input as random noise or start with a note/snippet of a piano sound, and do a probabilistic sampling of the next waveform sample. This next sample is fed back into the model to generate a sample again. We do not obtain meaningful sounding audio in this unconditioned setup, which can be due to several factors. Primarily, we do have a smaller context window of 100ms as compared to 240ms used in wavenet. It might also be a function of tuning temperature factor $T$, which also plays a major role in generating diverse sound rather than simple repetitions \cite{oord2016wavenet}.   Interestingly, for speech \cite{dieleman2018challenge} unconditioned models also fail to produce meaningful sounds. Most of the compelling results are conditioned on meta-data, timing, phonemes, etc. to name a few. However, similar to the theme in natural language processing as shown in \cite{kalchbrenner2016neural}, we outperform dilated convolutional-based models with transformer architectures for a next step prediction task for raw waveform generation as will be described in the next section. Some of the resulting audio examples can be found at:  https://tinyurl.com/6cjxffa8

\subsection{Transformer Implementation Details}

Heavy data augmentation was carried out like amplitude scaling, time shifts, etc. on the input data. The input one hot representations corresponding to the level of the bit representation was first passed through a positional encoding layer to add positional information before passing it to the Transformer architecture. For both the wavenet and Transformer architecture, the goal is to predict the next sample. Cross entropy loss was minimized between the ground-truth bit level and the predicted one over 256 levels. Adam optimizer \cite{kingma2014adam} was used with an initial learning rate of 1e-4 for the first 10 epochs, followed by 1e-5 and 1e-6 every time the loss values began to plateau. The training was performed on the Google Cloud Computing Platform for a maximum of 30 epochs for millions of data points. Dropout regularization was used for the feed-forward architecture to avoid over-fitting. For every training instance, instead of choosing to have a fixed dataset, we randomly shuffle training data at every epoch. The development and tuning were performed for a held-out validation set, and the best-performing models are evaluated. For every model, a batch size of 32 was chosen. For larger-scale experiments, we use NVIDIA V100 GPUs\footnote{The work is not necessarily predicated on the capacity of a specific GPU and can be run on slower GPUs as well, with smaller batch sizes. The only difference would be the time to train and iterate.}, and all of the frameworks were developed using open-source Tensorflow \cite{abadi2016tensorflow}, using distributed training. This will enable the work to be easily reproduced. For each of the specific architectures, the readers are advised to see Table 1 for further details for the number of attention heads, size of feed-forward architecture, and the embeddings.\footnote{All of the code was written using an open-source framework Tensorflow and can easily be reproduced.  There are several examples of Transformer implementations that can be tried from https://www.tensorflow.org/tutorials}

\section{Results and Discussion} 

For comparison against baseline neural generative models, we have chosen the prediction of the next sample as a criterion to compare different models. To quantify the performance, we choose to have top-5 accuracy as a metric. By this, we say that the model is accurate if one of the levels out of the top-5 predictions matches that of the ground truth. Such metrics have been used in error performance where there are many output categories e.g. \cite{gemmeke2017audio} \cite{deng2009imagenet}. For all the numbers reported in Table 1, we randomly subsample about 30min of data in total duration out of the 50 test tracks. No data augmentation or any other pre-processing was carried out. This is in contrast to what was reported in wavenet, which relied on Mean Opinion Scores (MOS) to quantify the performance of wavenet with that of other generative/audio synthesis algorithms. We reason that the output of a system such as a text-to-speech system is dependent on several variables. For example, how well can a model handle external attributes/conditioning, the performance of spectral synthesis over waveform synthesis, etc. There can also be errors introduced due to the generation of waveforms from spectral representations too. Additionally, there can be different errors introduced due to different applications e.g. a mean opinion score metric for the TTS system might not be a good metric for other perceptual tasks such as speech denoising/enhancement. Finally, concerning the quality of generated audio in unconditioned setup without meta-data, it again is a function of how to sample the predictions, how we modify the temperature function. We evaluate it against the wavenet model as it is state-of-the-art in raw waveform synthesis.

Due to these reasons, we focus on the synthesis of waveform alone and report raw accuracy numbers.  Since for both wavenet, the objective for training is how good the model is in the next step prediction using cross-entropy loss, we believed that instead of giving negative log-likelihood scores, we give the accuracy as to how each of them does in the objective criteria they are trained on. Table 1 compares the performance of all the systems tried.  For the vanilla wavenet system, we fix the hyper-parameters according to the one implemented by Google in \cite{oord2016wavenet}. The vanilla wavenet architecture consists of 10 layers of causal convolutions with each layer having 128 filters, and the receptive field was dilated by a factor of 2 in each layer with the initial filter size being 2. The second variant of this model stacked 3 such modules yielding a total of 30 layers. We see that with depth the model achieves a performance of about 76\% as opposed to 74\% with the vanilla model. This is because the stacked model learns even longer context as opposed to the vanilla architecture. We did not tune on the filter sizes, the number of filters, and these parameters were chosen from the paper as described in \cite{oord2016wavenet}. 

For the Transformer architecture, we first choose the parameters to have 4 attention heads, with the size of latent space being 128, and the size of the feed-forward architecture was chosen to be 256. For most of the Transformer implementation seen across literature  \cite{dosovitskiy2020image} typically the number of attention heads are chosen to be either 4 or 8 and we adhered to such a choice. We see that the network already outperforms even the large-scale 30-Layer wavenet architecture by a significant margin of about 4\%. One of the main reasons we attribute the success of these models is because attention models learn which parts of the waveform are important. This enables learning a much more complex topology than the causal convolutional filters learned by wavenet architectures. Additionally, when we condition it on the past waveform, it outperforms the same architecture without conditioning by 2 \%. Even though this is a small number, it is as significant of a jump as between a Vanilla wavenet implementation and a Stacked 30-Layer wavenet model. One of the hypotheses as to why this happens is the fact that we are in some manner able to capture even longer contexts by conditioning the generation on past samples. For example, the latent code can learn pitch patterns, chord distributions, which can guide the output predictions better. In the future, speaker/instrument-specific conditioning can be carried out to generate a diverse range of audio signals. This flexibility of conditioning via convolutional architectures  can be modified in future for a variety of applications e.g. Packet Loss Concealment \cite{verma2020deep,rottondi2016overview}, instrument transfer \cite{haque2018conditional,mor2018universal} to name a few. Finally, to see how much we can improve the performance with large deeper Transformer architectures, we train an 8-Layer Transformer model, with 8 attention heads keeping the size of embeddings and the architecture of feed-forward architecture the same. We see that the performance increases significantly yielding 5\% absolute improvement over the baseline Transformer architecture. This is in line with what has been seen in the literature of improved performances with the size of the models\cite{fedus_zoph_shazeer_2021}. The improvement over traditional wavenet architectures with comparable architecture by such significant margins is exciting for the future of raw audio synthesis. 

\begin{table}[ht]
  \caption{\itshape Comparison of various proposed architecture as shown in the table below for top-5 prediction accuracy. H: The number of attention heads used in our model, E: the size of the embedding at each layer of Transformer. d: dilation rate at every level for wavenet architecture, N: Number of Layers, F: Number of filters in each layer. The Transformers here refer to Transformer architecture with Causal Attention where prediction at each time step depends on previously seen values.}
	\centering
	\begin{tabular}{|c|c|}
		\hline
		Neural Model Architecture & Accuracy \\\hline
		Vanilla Wavenet: d = 2, N = 10, F = 128 & $74\%$ \\
		Stacked Wavenet: d = 2, N = 30, F = 128 & $76 \%$ \\\hline
		\\3-Layer Transformer: H = 4, E = 128 & $80 \%$ \\
		Conditioned 3-Layer Transformer: H = 4, E = 128 & $82 \%$ \\
		Large 6-Layer Transformer: H = 8, E = 128 & $84 \%$ \\
		Large 8-Layer Transformer: H = 8, E = 128 & $85 \%$ \\\hline
	\end{tabular}
	\label{tab:example}
\end{table}

\section{Conclusion}
With the advent of the success of generative Transformer based models in a variety of domains, this paper presents how they have outperformed state-of-the-art architecture for raw audio synthesis. We show how using conditioned generative models, we outperform a classical wavenet architecture for raw audio synthesis on the next step prediction task. By combining classical latent representation with that of a Transformer architecture, we show we can incorporate even longer contexts with current architectures. This has the potential to improve the classical pipelines where a wavenet-based generator was used e.g. audio synthesis, source separation, denoising, audio transforms, vocoders to name a few. It is flexible to incorporate conditioning on external attributes, as in our case, the past context. 
\section{Future Work}
Several future research directions can be explored with the current work. It will be interesting to compare the performance of waveform-based generative models for applications such as TTS, audio source separation, denoising, packet loss concealment.  Another future direction would be to use variants of a standard transformer architecture to improve the performance further e.g. sparse transformers, switch transformers. It would also be interesting to see what the Transformer architecture learns in the attention maps, and better interpret them. We kept getting improved performance as we increased the model complexity, which makes the prospect of further increasing the depth and size of the Transformer module for improved performance. As future work, we want to explore synthesis by giving more context than 100ms, which will make the unconditioned generation of audio possible as shown in wavenet work. Further, we want to push the envelope of how far we can get, both in terms of modeling longer-term dependencies of the past samples over several seconds on a waveform level to pave way for more exciting applications.

\section{Acknowledgements} This work was done as part of a grant from the AHRC in the UK entitled "Hybrid Live". We thank our partners at SF Jazz Center. This work was also supported by the Institute of Human-Centered AI at Stanford University (Stanford HAI) through a generous Google Cloud computing grant.  It made training and iterating over these massive networks possible, and have a level playing field in terms of computing resources, to a certain extent, with industry researchers. We also thank our DAFX reviewers and Camille Noufi for their feedback on the manuscript.

\nocite{*}
\bibliographystyle{IEEEbib}
\bibliography{DAFx20_tmpl} 

\end{document}